# To Change Or To Stick: Unveiling The Consistency Of Cyber Criminal Signatures Through Statistical Analysis

Ronan Mouchoux, University of South Brittany
François Moerman, XRATOR

**Abstract.** This study unveils the elusive presence of criminal signatures in cyberspace, validating for the first time their existence through statistical evidence. By applying the A priori algorithm to the modus operandi of Advanced Persistent Threats, extracted from an extensive corpus of over 17,000 articles spanning 2007 to 2020, we highlight the enduring patterns leveraged by sophisticated cyber criminals. Our findings verify the existence of unique signatures associated with advanced cybercriminals, bridging a crucial gap in current understanding of human behavior in cyber-attacks. This pivotal research sets the foundation for an entirely new academic intersection in cybersecurity and computational criminology.

> *"We are what we repeatedly do. Excellence, then, is not an act, but a habit."*
> The Story of Philosophy, Will Durant

> *"The great art is to change during battle. Woe betide anyone who comes into battle with a battle plan definitive."*
> Military Maxims of Napoleon, Napoléon Bonaparte

## 1. Introduction

The nature of the digital battlefield is such that patterns of behavior, repeated tactics, and distinct modus operandi can often unveil the hidden signatures of cybercriminals. However, unlike traditional criminal investigations where physical evidence might provide direct links to the perpetrators, the intangible nature of cybercrimes often masks these telltale signs, making them elusive to detection.

Durant's philosophy suggests that our identity is molded by our habits, hinting at the possibility of uncovering unique signatures based on repeated actions [1]. In the cybersecurity context, this concept translates into exploring the persistent operational patterns of cybercriminals that could potentially serve as their unique 'digital signatures.' At the same time, Bonaparte's maxim underlines the importance of adaptability and change during conflict, a principle that rings true in the ever-changing cybersecurity landscape [2]. The dynamic nature of cyber confrontation necessitates adaptability and continuous evolution in cyber defense strategies and in offensive operation.

Our study seeks to identify the existence of criminal signatures in the realm of cyberspace, specifically focusing on Advanced Persistent Threats (APTs). By analyzing an extensive corpus of over 17,000 articles spanning from 2007 to 2020 using the A priori algorithm, we aim to illuminate the enduring patterns of these sophisticated cybercriminals. This corpus is the result of a previous research made by the authors [3]. The objective of this work is twofold: to provide a robust statistical validation of the existence of criminal signatures in cyberspace, and to establish a new academic intersection in cybersecurity and computational criminology, that we called Computational Cyber Threat Intelligence.

Through this research, we intend to bridge the crucial gap in understanding human behavior in cyber-attacks, which could significantly enhance our predictive capabilities and contribute to the development of more effective cyber defense strategies. In the following sections, we will detail our research methodology, present our findings, and discuss the implications of our work for both academia and practical applications in the cybersecurity sector.

## 2. Theorical Background

Modus operandi analysis, rooted in the Routine Activity Theory in criminology, helps identify habitual patterns in criminal behavior, and the same principles apply to the cyber domain [4][5][6][7][8]. Investigators' interpretations in forensic analysis, including cyber forensics, are abductive and inductive, limited by partial knowledge of the event chain and their environment [9].

On the industrial side, Cyber Threat Intelligence (CTI) applies these concepts, characterizing cyber-attack methods and tools as Tactics, Techniques, and Procedures (TTP)[10]. Models such as the "Pyramid of Pain" and "Detection Maturity Level Model" advocate modus operandi analysis for enhancing cybercrime detection and threat actor deterrence and have been widely adopted by practitioners, despite their lack of academic foundation [11][12]. The MITRE ATT&CK® Framework, along with OASIS STIX, presents a widely accepted lexicon and format for describing cyber threat activity [13][14].

Association Rules Mining like apriori can augment knowledge and uncover hidden technique patterns, improving cyberattack attribution and with the potential to assist the threat hunting process [15]. Yet, in research literature, when it comes to CTI, the apriori algorithm seemed mainly applied to technical artefact such as IP address, email, or URL [16][17].

Zhi Li et al.'s study leverages the Apriori algorithm to analyze global cyberspace security issues [18]. By looking at over 56,000 web pages from diverse professional and non-professional websites, they identify key cybersecurity topics and their correlations. This research successfully applied A priori algorithm to the cybersecurity domain literature, but do discover relationship and correlation between topical issues, and not adversaries modus operandi descriptors.

Attempts to identify criminal patterns with computational techniques exist in traditional criminology, but a gap exists in applying these techniques to cyber domains and addressing cyber-specific peculiarities [19][20]. Traditional pattern discovery methods applied to crime data tend to overlook some relationship, when a priori effectively reveals detailed patterns and correlations, thereby enable the creation of actionable rules for crime prevention and detection [21].

Furthermore, studies integrating behavioral psychology into cybersecurity primarily focus on potential victims' behavior rather than the attackers [22][23][24][25][26].

This study argues that cyber threat actors develop habits and preferences in their TTPs, constrained by the attacker's individual and organizational comfort zones and the targeted environment. Even if we can only detect and document what threat actors allow us to see, we postulate that it is more challenging for threat actors to change their human habits (reflected in TTPs) than the technical traces left by their tools and infrastructures. Hence, this research introduces the AbductionReductor tool that applies the A priori algorithm to threat actors' technique sets to validate these convictions. The study represents the first attempt to profile cyber threat actors using this method, providing quantitative and qualitative analyses of technique association rules, and underscoring the existence of a criminal signature in cyberspace.

## 3. Data Collection

The data for this study was curated and structured based on the authors' prior work [3]. A total of 17153 articles from 26 publicly accessible sources covering a period from 2007-10-27 to 2020-09-22, were assembled to investigate the TTPs used by various cyber threat actors.

The collected data was automatically structured in accordance with the MITRE ATT&CK® framework (v6.3), providing a universally recognized taxonomy and ensuring a consistent format. This systematic organization facilitated the efficient application of the Apriori algorithm for pattern discovery and association rule mining.

In addition, the paper utilized a malware reference base composed of 2790 malwares names and aliases, a threat actors reference base composed of 851 adversaries' names and aliases, a techniques reference base composed of 535 techniques denomination and synonyms, and a location reference base composed of 716 places. These references were used to help classify and organize the data collected from the blog articles.

The resulting data was stored in an Entity Database (Entity DB). It contains the category of the entity (CVE, Location, Malware, Technique, Threat Actor), the matching phrase, the matching sequence, the matched entity, the reference of the source article. By extracting entities from the textual data, and applying techniques like disambiguation and name resolution, the resulting Entity Database is more consistent and reliable for analysis.

For this study, we used two data structures. The first one focus on the technique (STIX Attack Pattern Domain Object). The second one focus on the technique, the malware and the vulnerability, a grouping referred as an Intrusion Set (STIX Intrusion Set Domain Object). We expose in this paper the result for the first data structure.

The first data structure used for association rule mining in the study was a transactional database format, where each transaction represents a document in the Knowledge Database, and each item in the transaction represents a technique extracted from that document.

The threat actor is a crucial feature for clustering several documents, while the techniques indicate the methods employed by the attacker to achieve their goals. These features are used to generate association rules, which are patterns of co-occurring techniques that are frequently observed in the analyzed data.

## 4. Association Rule Mining

AbductionReductor is an association rules mining tool developed with cyber-attack investigation in mind. We developed it based on the observation that cyber-attack investigation is an abductive process, where an investigator must induce the cause of observed effects from technical traces while having a partial vision of the incident or the global adversary's activity. In addition, our hypothesis is that threat actors tend to use the same techniques repeatedly, which aligns with the "habitual criminal" concept in criminology, creating technical "comfort zone" which it is hard and costly for an individual or an organization to deviate.

AbductionReductor is designed to extract useful information from large datasets of cyber-attack reports, specifically focusing on the techniques used by threat actors. The tool employs association rule mining to identify patterns in the data, generating rules in the form of "*If technique X is present (antecedent technique), then technique Y is also*

*likely to be present (consequent technique).*" These rules describe a relationship but does not imply causality. The qualitative interpretation of the relationship will be performed during the post analysis process, based on Cyber Threat Intelligence and Offensive Security domain knowledge.

In our implementation, there can be only one item in the antecedent technique set and only one item in the consequent technique set. This is designed to improve the interpretability of the result.

We use four parameters for our association rule mining: Minimal candidate support, Antecedent technique support, Association confidence, Lift metric. We will primarily base our qualitative analysis on threat actors with the most association rules, then on association with the highest lift metrics.

**Table 1.** Association rules mining parameters

| Parameter | |
|---|---|
| Candidate Support (CS) | 3 + 2% |
| Antecedent Support (AS) | 2 * CS |
| Association Confidence (AC) | 50% |

In the data structure, 12 808 articles (transactions) qualified to be in the transaction table. To be qualified, an article must have a single identified threat actor (exclusion of multi-actor articles) and must have at least two techniques. The result is 6 353 associations rules spread over 73 unique threat actors (dataset n°1). To focus on highly correlated association, we further filter on the lift metric (equal or greater than 1) we have 901 association rules spread over 16 unique Threat Actors (dataset n°2).

**Table 2.** Measures on parameters of the technique association rule table (dataset n°2)

| Measures | Antecedent Support (AS) | Association Confidence (AC) | Lift Metric |
|---|---|---|---|
| Minimal value | 6 | 50 | 1 |
| Maximal Value | 64 | 100 | 5.440 |
| Mean | 19.599 | 69.792 | 1.597 |
| Median | 15 | 66 | 1.400 |
| Mode | 8 | 66 | 1.070 |
| Standard deviation | 13.030 | 16.483 | 0.646 |
| Interquartile range (IQR) | 15 | 18 | 0.590 |

**Table 3.** Top five Threat Actors by number of rules (dataset n°2)

| Measures | Nbr. Rules | Mean Support (Standard Deviation) | Mean Confidence (Standard Deviation) | Lift (Standard Deviation) |
|---|---|---|---|---|
| APT28 | 180 | 27.728 (16.8) | 72.183 (17.1) | 1.577 (0.71) |
| ELECTRUM | 114 | 21.333 (11.9) | 67.632 (16.7) | 1.731 (0.73) |
| EQUATION | 102 | 20.922 (12.8) | 67.284 (16.9) | 1.592 (0.56) |
| COVELLITE | 97 | 12.495 (6.7) | 67.426 (16.6) | 1.812 (0.71) |
| TURLA | 97 | 16.340 (8.8) | 69.526 (15.6) | 1.615 (0.67) |

## 5. Threat Actor Profiling

Using the results obtained from association rule mining, we perform a threat actor profiling, focusing of the rules with highest lift, and reviewing qualitatively the top associations for each selected threat actors. We finally explore the similarities and differences between the TTPs of the selected threat actors and discusses the implications of these findings for cybersecurity.

### a. Selection of the Threat Actor

In the dataset n°2 we have 16 threat actors. We select the top 5 in terms of number of rules, as it should provide more robust association rules metrics.

The threat actors' names we used are the result of the Threat Actor Name Disambiguation and Resolver of the preprocessing step. Behind each name lays several aliases.

**Table 4.** Selected Threat Actors names and Aliases

| Name | Aliases |
|---|---|
| APT28 | apt28, snakemackerel, swallowtail, group 74, sednit, sofacy, pawn storm, fancy bear, strontium, tsar team, threat group-4127, tg-4127, apt 28, pawnstorm, tsarteam, group-4127, tag_0700, iron twilight, sig40, apt_sofacy |
| ELECTRUM | sandworm team, sandworm, electrum, black energy, blackenergy, temp.noble, iron viking |
| EQUATION | equation, tilded team, equation group, lamberts, longhorn, eqgrp, the lamberts, apt-c-39 |
| TURLA | turla, waterbug, whitebear, venomous bear, snake, krypton, turla group, group 88, wraith, turla team, uroburos, pfinet, tag_0530, pacifier apt, popeye, sig23, iron hunter, makersmark, skipper turla, white bear |
| COVELLITE | lazarus group, hidden cobra, guardians of peace, zinc, nickel academy, andariel, operation darkseoul, dark seoul, hastati group, unit 121, bureau 121, newromanic cyber army team, bluenoroff, group 77, labyrinth chollima, operation ghostsecret, operation applejeus, whois hacking team, appleworm, apt-c-26, lazarus, covellite, |

### b. Threat Actor Based on their Association Rules

For each of the five threat actors, we perform a selection of their top association rules based on the top 5 highest lift metrics. We qualitatively review each technique association from an Offensive Security or Cyber Threat Intelligence perspective.

### i. APT28

APT28 is suspected to be a state-sponsored sophisticated group. It is suspected to work for the same organization as ELECTRUM. It is also suspected to work for the same state as TURLA.

**Table 5. APT28's top 5 technique associations by lift metrics**

| ID | Antecedent Technique | Consequent Technique | Support | Confidence | Lift |
|---|---|---|---|---|---|
| T1 | Custom Cryptographic Protocol (T1024) | Logon Scripts (T1037) | 12 | 100 | 5,44 |
| T2 | Modify Registry (T1112) | Registry Run Keys / Startup Folder (T1060) | 8 | 80 | 5,23 |
| T3 | Rundll32 (T1085) | Logon Scripts (T1037) | 8 | 75 | 4,59 |
| T4 | Software Packing (T1045) | Windows Management Instrumentation (T1047) | 8 | 75 | 4,59 |
| T5 | Process Discovery (T1057) | Peripheral Device Discovery (T1120) | 20 | 100 | 3,5 |

**Table 6. APT28's top 5 technique associations qualitative review**

| ID | Observation |
|---|---|
| T1 | Persistence with antivirus evasion (packing). |
| T2 | Two-stage persistence: Startup folder and Run keys are highly scrutinize by endpoint protection software, so the attacker will put legitimized executable (such as signed executable) that will load a payload stored as a registry value. |
| T3 | This is an implementation of the first stage of a two-stage persistence operation, where rundll32 can be a run key or an Active Directory logon script. |
| T4 | Internal delivery via WMI of a packed malicious payload to avoid detection, for privileged escalation or persistence. |
| T5 | Common task of a host reconnaissance and monitoring operation. |

### ii. ELECTRUM

ELECTRUM is suspected to be a state-sponsored sophisticated group. It is suspected to work for the same organization as APT28. It is also suspected to work for the same state as TURLA.

**Table 7. ELECTRUM's top 5 technique associations by lift metrics**

| ID | Antecedent Technique | Consequent Technique | Support | Confidence | Lift |
|---|---|---|---|---|---|
| T1 | Drive-by Compromise (T1189) | Man in the Browser (T1185) | 8 | 75 | 4.59 |
| T2 | Remote Access Tools (T1219) | Input Capture (T1417) | 8 | 75 | 4.59 |
| T3 | Clipboard Data (T1115) | Input Capture (T1417) | 8 | 75 | 4.59 |

| ID | Antecedent Technique | Consequent Technique | Support | Confidence | Lift |
|----|----------------------|----------------------|---------|------------|------|
| T4 | External Remote Services (T1133) | Remote Services (T1021) | 10 | 75 | 3.94 |
| T5 | System Information Discovery (T1426) | System Firmware (T1019) | 10 | 75 | 3.94 |

Table 8. ELECTRUM's top 5 technique associations qualitative review

| ID | Observation |
|----|-------------|
| T1 | Watering hole or communication interception and payload injection (PRISM-like) |
| T2 | Keylogger |
| T3 | Monitor some input (like CTRL+C) to trigger the inspection of the content of the clipboard. |
| T4 | Using valid account on a VPN, RDP, or external accessible services. |
| T5 | Information gathering for persistence or privilege escalation (Rootkit-like). |

### iii. EQUATION

EQUATION is suspected to be a state-sponsored sophisticated group. It is believed to be working for a state ideologically opposed the one of APT28 or the one of COVELLITE.

Table 9. EQUATION's top 5 technique associations by lift metrics

| ID | Antecedent Technique | Consequent Technique | Support | Confidence | Lift |
|----|----------------------|----------------------|---------|------------|------|
| T1 | Input Capture (T1417) | Clipboard Data (T1115) | 17 | 100 | 3.74 |
| T2 | Remote Access Tools (T1219) | File and Directory Discovery (T1420) | 11 | 80 | 3.59 |
| T3 | Credential Dumping (T1003) | Clipboard Data (T1115) | 13 | 80 | 3.19 |
| T4 | Software Packing (T1045) | Execution through Module Load (T1129) | 11 | 75 | 3.16 |
| T5 | Execution through API (T1106) | Execution through Module Load (T1129) | 22 | 100 | 2.99 |

Table 10. EQUATION's top 5 technique associations qualitative review

| ID | Observation |
|----|-------------|
| T1 | Same as ELECTRUM-T3. |
| T2 | The attacker is using the RAT to performs its host reconnaissance. |
| T3 | Stole credential that pass through the clipboard. |
| T4 | First phase of a two-stage persistence operation. |
| T5 | Defense evasion to load external payload using a low-level Windows API. |

### iv. TURLA

TURLA is suspected to be a state-sponsored sophisticated group. It is suspected to work for the same state as APT28 and ELECTRUM, but for a different organization.

**Table 11. TURLA's top 5 technique associations by lift metrics**

| ID | Antecedent Technique | Consequent Technique | Support | Confidence | Lift |
|---|---|---|---|---|---|
| T1 | Data Compressed (T1002) | Remote Access Tools (T1219) | 6 | 66 | 3.41 |
| T2 | Remote Access Tools (T1219) | Data Destruction (T1485) | 6 | 66 | 3.41 |
| T3 | Remote Access Tools (T1219) | Data Compressed (T1002) | 6 | 66 | 3.41 |
| T4 | Data Destruction (T1485) | Remote Access Tools (T1219) | 6 | 66 | 3.41 |
| T5 | Data Encoding (T1132) | Windows Management Instrumentation (T1047) | 14 | 100 | 3.29 |

**Table 12. TURLA's top 5 technique associations qualitative review**

| ID | Observation |
|---|---|
| T1 | Data exfiltration. |
| T2 | Turla's malwares generally includes the ability to delete files (e.g.: CompFun, Reductor, TwoFace web shell). |
| T3 | Same as TURLA-T1. |
| T4 | Same as TURLA-T2. |
| T5 | Avoid network inspection during data exfiltration or lateral movement. |

### v. COVELLITE

COVELLITE is suspected to be a state-sponsored sophisticated group. Its suspected state is opposed ideologically to EQUATION's one and compatible with APT28's one.

**Table 13. COVELLITE's top 5 technique associations by lift metrics**

| ID | Antecedent Technique | Consequent Technique | Support | Confidence | Lift |
|---|---|---|---|---|---|
| T1 | Data Encoding (T1132) | Data Encrypted (T1022) | 8 | 100 | 5.29 |
| T2 | Data Encrypted (T1022) | Data Encoding (T1132) | 6 | 75 | 3.97 |
| T3 | Custom Cryptographic Protocol (T1024) | Data Encrypted (T1022) | 13 | 100 | 3.53 |
| T4 | System Time Discovery (T1124) | Data Encrypted (T1022) | 13 | 100 | 3.53 |
| T5 | Process Hollowing (T1093) | Hidden Files and Directories (T1158) | 6 | 66 | 3.13 |

**Table 14. COVELLITE's top 5 technique associations qualitative review**

| ID | Observation |
|---|---|
| T1 | Network inspection evasion. |
| T2 | Same as COVELLITE-T1. |
| T3 | Exfiltration of encrypted data with a custom encryption algorithm. |
| T4 | Time-based spacing exfiltration. |
| T5 | Two stage persistence where the payload to inject in the hollowed process is store as a hidden file or in a hidden directory. |

### c. Commonalities and Differences Between TTPs

We combine all association rules among all selected threat actors with a lift greater than 1. We have a total of 590 association rules including 445 unique associations (75%). The distribution is the following:
- APT28: 180
- ELECTRUM: 114
- EQUATION: 102
- TURLA: 97
- COVELLITE: 97
- TOTAL: 590

Computing the most shared association rules across threat actors, we can notice that they concern mainly techniques relative to furtivity tactics and reconnaissance of the victim's internal network tactics:
- T1140-T1045: Deobfuscate/Decode Files or Information => Software Packing
- T1027-T1140: Obfuscated Files or Information => Deobfuscate/Decode Files or Information
- T1045-T1140: Deobfuscate/Decode Files or Information => Deobfuscate/Decode Files or Information
- T1426-T1057: System Information Discovery => Process Discovery
- T1027-T1002: Obfuscated Files or Information => Data Encrypted

**Table 15. Top 5 technique association**

| Association | Count | Threat Actors |
|---|---|---|
| T1140-T1045 | 4 | APT28, EQUATION, TURLA, COVELLITE |
| T1426-T1057 | 4 | APT28, ELECTRUM, TURLA, COVELLITE |
| T1027-T1140 | 3 | APT28, ELECTRUM, EQUATION |
| T1045-T1140 | 3 | EQUATION, TURLA, COVELLITE |
| T1027-T1002 | 3 | APT28, ELECTRUM, EQUATION |

In apriori, the association is directed, meaning that A-B is not the same as B-A. We use the Jaccard similarity coefficient (intersection divided by union) and the Sørensen–Dice coefficient (intersection divided by the size of the smaller set).

Table 16. Threat Actor's technique association rules similarity (Jaccard / Dice)

| Threat Actor | ELECTRUM | EQUATION | TURLA | COVELLITE |
|---|---|---|---|---|
| APT28 | 0.083/0.175 | 0.175/0.117 | 0.155/0.083 | 0.125/0.155 |
| ELECTRUM | x | 0.052/0.088 | 0.062/0.051 | 0.093/0.072 |
| EQUATION | x | x | 0.052/0.088 | 0.072/0.062 |
| TURLA | x | x | x | 0.125/0.093 |

We interpret those metrics as showing very little overlap between threat actors' technique association rules. As we have seen that our association rules listing is composed at 75% of unique combination, we perform a second similarity comparison between actors with association that appears at least among two threat actors' sets.

We combine all association rules, that appear at least twice among different threat actors' sets, with a lift greater than 1, among all selected threat actors. We have a total of 125 association rules. The distribution is the following:
- APT28: 45
- ELECTRUM: 33
- EQUATION: 23
- TURLA: 19
- COVELLITE: 28
- TOTAL: 125

Table 17. Threat Actor's repetitive technique association rules similarity (Jaccard / Dice)

| Threat Actor | ELECTRUM | EQUATION | TURLA | COVELLITE |
|---|---|---|---|---|
| APT28 | 0.421/0.060 | 0.600/0.522 | 0.522/0.421 | 0.606/0.536 |
| ELECTRUM | x | 0.263/0.391 | 0.261/0.263 | 0.473/0.250 |
| EQUATION | x | x | 0.400/0.263 | 0.391/0.261 |
| TURLA | x | x | x | 0.263/0.474 |

Even by focusing on repetitive techniques, we interpret those metrics as showing little similarities between threat actors' technique association rules.

We can also acknowledge that the distribution of association rules is not uniform among the different threat actors, with APT28 having the highest number of association rules and COVELLITE having the lowest, by 2-fold for the full rules set and for the repetitive sets. This impact necessarily the results and may impact their interpretation.

### d. Implication for Cyber Security

The identification of common techniques used by different threat actors emphasizes the criticality of obfuscation for defense evasion, and internal reconnaissance and monitoring for cyber operation. It may also indicate that, at the technique level, their may not be a lot of choice for the attackers of how to perform those tasks. It may also indicate that the underlying normalization vocabulary (MITRE ATT&CKv6.3) is not enough granular for these types of tactics and techniques.

The low overlap in technique association rules among sophisticated threat actors, suggesting unique modus operandi. This confirms the need for cyber threat intelligence analysts to grasp the distinct traits of each threat actor for effective identification and response. The methodology outlined here could guide analysts on what aspects to scrutinize and where to concentrate their efforts.

Association rules can also assist incident response investigation by uncovering relationships between techniques commonly used in pairs. This can help prioritize investigative efforts and potentially identify additional indicators of attacks that may have been missed otherwise.

In any case, association rules should not be used in isolation and should be combined with other techniques and approaches for a comprehensive incident response plan. They must also be interpreted in the context of the incident by a domain specialist.

## 6. Discussion

This research presents a significant leap in the field of Cyber Threat Intelligence and advances our understanding of criminal modus operandi in the cyber domain. We harnessed our collective expertise in CTI, criminology, forensics, and offensive security to construct a novel methodology that builds upon the traditional Apriori association rules mining algorithm and the traditional criminal profiling.

Our approach emphasizes the unique behavioral characteristics of threat actors. Cyber attackers, like traditional criminals, exhibit habitual behaviors - or preferences - which are reflected in their Tactics, Techniques, and Procedures (TTPs). We introduced previously the concept of the "Cyber Operation Constraint Principle" to encapsulate this idea [3]. In essence, we postulate that while threat actors adapt their TTPs to achieve specific intrusions, their choices are influenced by their own habits and preferences, as well as the constraints of the targeted environment. The result of this work is to move this assertion from a postulate to empirical evidence: our belief, that even at a higher level of abstraction (the technique level) than forensic artefact we can still discriminate threat actors based on their technique association rules, is statistically verified by the experiment.

Yet, the methodology currently processes all TTPs within a single threat actor's set without considering the temporal factor. This approach doesn't allow observing evolution or adherence to specific technique associations over time. Moreover, potential inaccuracies during data collection and extraction may affect the results. However, we remain convinced that these limitations can be addressed with future research and improvements in our methodology using state-of-the-art techniques [27]. Finally, as we are industry practitioners and not academic researchers, this study probably has flaws, such as an incomplete literature review, a custom methodology and a superficial interpretation and discussion of the results and implication of our experiment.

We plan to improve the data collection and preprocessing stages and aim to create multi-level directed graphs that consider technique associations in accordance with time and higher-level intrusion phases. Our vision is to build a comprehensive understanding of how threat actors operate, evolve, and adapt their TTPs.

## 7. Conclusion

This study marks a significant evolution in the intersection of Cyber Threat Intelligence, computational criminology, and behavioral psychology. We called it Computational Cyber Threat Intelligence. It sheds light on the modus operandi of sophisticated threat actors, revealing the unique 'digital signatures' that these cybercriminals leave behind, despite the intangible and highly adaptive nature of cybercrime.

Using association rule mining and the insights from cyber-criminal profiling, we've demonstrated that specific, distinguishable patterns of behavior do indeed exist in the cyber domain. These findings echo Durant's philosophy [1] that our identity is in part sculpted by our habits, and they hold true even in the realm of advanced cyber threats. We have moved the threat actor modus operandi topics from hypothesis and industry practitioners' informal models into a landscape of tangible, statistical empirical evidence that supports the existence of unique threat actor habits and preferences.

Our research, however, is not without its limitations. False positives in the entity extraction process and the potential for data poisoning during the collection of threat actors' attacks are areas that need to be improved. We also acknowledge the limitation of our current approach in its consideration of temporal factors that could provide critical insights into the evolution of threat actors' behaviors over time. We also warned the readers that we are industry practitioners and not academic researchers, and that our study may have methodological flaws.

Despite these challenges, we see a promising direction for future work. We intend to refine our data collection and preprocessing methodologies and incorporate time references and higher-level intrusion phase representations into our technique association rules analysis. This advancement will provide a more nuanced understanding of how cyber threat actors operate and adapt their TTPs over time, leading to more robust and effective defensive strategies.

Our research has revealed the potential of a data-driven approach to understanding and predicting cybercriminal behavior. We have initiated an original dialogue at the intersection of cybersecurity, criminology, and behavioral psychology, and we believe that this work is only the beginning. As we continue to delve deeper into this critical field, we are hopeful that our efforts will significantly contribute to bridging the gap in cybersecurity defenses and lead to a safer and more secure digital world by sharpening the understanding of criminal behavior.